\documentclass[prb,twocolumn,aps,showpacs]{revtex4}
\usepackage{graphicx}
\usepackage{bm}

\begin{document}
\date{\today}

\title{Comparison between a diagrammatic theory for the BCS-BEC 
crossover and Quantum Monte Carlo results}
\author{P. Pieri, L. Pisani, and G.C. Strinati}
\affiliation{Dipartimento di Fisica, 
Universit\`{a} di Camerino, I-62032 Camerino, Italy}

\begin{abstract}
Predictions for the chemical potential and the excitation gap recently obtained
by our diagrammatic theory for the BCS-BEC crossover in the superfluid phase 
are compared with novel Quantum Monte Carlo results at 
zero temperature now available in the literature. A remarkable agreement 
is found between the results obtained by the two
approaches.
\end{abstract}


\maketitle
The recent experimental realization of the BCS-BEC crossover with ultracold
trapped Fermi atoms\cite{exp-BCS-BEC} has given impetus to  
theoretical investigations of this crossover. 
In a recent paper\cite{PPS-PRB-04}, the t-matrix self-energy approach 
(originally conceived for the
normal phase\cite{Perali02}) was extended to the superfluid phase, aiming at 
improving the description of the BCS-BEC crossover by including pairing 
fluctuations on top of the BCS
mean-field approach considered in Refs.~\onlinecite{Leggett} and 
\onlinecite{NSR}.

In this theory, the effects of the collective Bogoliubov-Anderson mode is
explicitly included in the fermionic self-energy, thus generalizing  
the theory due to Popov for a weakly-interacting (dilute) superfluid Fermi 
gas\cite{Popov}.
The theory is based on a judicious choice of the fermionic self-energy, 
such that it reproduces the fermionic mean-field BCS behavior plus pairing 
fluctuations in the weak-coupling limit as well as the Bogoliubov description
for the composite bosons which form in the strong-coupling limit.
In the intermediate-coupling region of interest about the unitarity limit,
where no small parameter exists to control the many-body approximations, the
theory is able to capture the 
essential physics of the problem, as the excellent agreement with a previously
available QMC calculation\cite{carlson03} at the unitarity point 
$(k_F a_F)^{-1}=0$ has already shown\cite{PPPS-PRL-04}, and as more extensively
demonstrated by the present comparison with more recent QMC 
data\cite{chang04,FNQMC} spanning the
whole crossover region. The theory of Ref.~\onlinecite{PPS-PRB-04} is 
completely {\em ab initio\/} and it contains no adjustable parameter. Although
the comparison with QMC data is here limited to the zero-temperature limit 
where they are available, the predictions of the theory of 
Ref.~\onlinecite{PPS-PRB-04} extend  as well to finite temperature 
and across the critical temperature. 

Purpose of this Brief Report is to compare the theoretical predictions 
obtained 
from the theory of Ref.~\onlinecite{PPS-PRB-04} with novel 
Quantum Monte Carlo (QMC) 
data\cite{chang04,FNQMC}, which were published after completion of
Ref.~\onlinecite{PPS-PRB-04}. A quantitative comparison between the results 
for the density 
profiles obtained from a local density version\cite{PPPS-PRL-04} to the 
theory of Ref.~\onlinecite{PPS-PRB-04} and the experimental data was already 
presented in Ref.~\onlinecite{quantitative}.

Both our calculations and the QMC calculations of Refs.~\onlinecite{chang04} 
and \onlinecite{FNQMC} are based on a model Hamiltonian describing a system of 
fermions mutually interacting via an attractive contact potential. In 
Ref.~\onlinecite{SPS} this 
Hamiltonian was proved  appropriate 
to describe the BCS-BEC crossover with trapped Fermi gases.
In this model, 
the only dimensionless parameter representing the effective coupling strength
is the (inverse of the) product $k_F a_F$ between the Fermi wave vector
$k_F$ and the fermionic scattering length $a_F$. For the homogeneous gas here 
considered,
$k_F=(3 \pi^2 n)^{1/3}$ where $n$ is the particle density.
Comparison will be made at zero temperature only, since 
finite-temperature QMC calculations for the BCS-BEC crossover are not yet 
available.
 
The overall agreement between the two alternative (diagrammatic and QMC) 
calculations turns out to be quite good, 
expecially in the most interesting intermediate-coupling regime about 
$(k_F a_F)^{-1}=0$.
\begin{figure}
\begin{center}
\includegraphics*[width=8.5cm]{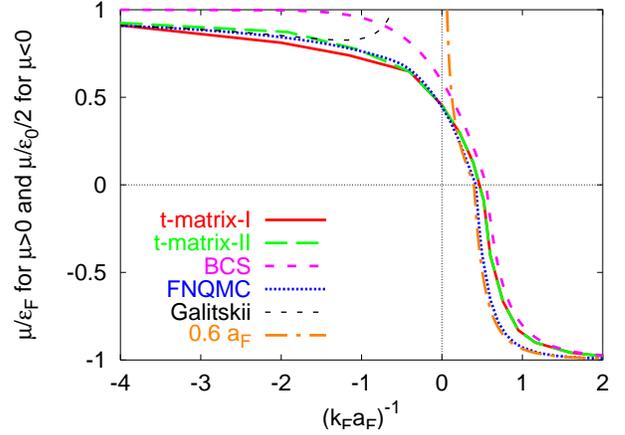}
\caption{Chemical potential at zero temperature vs the coupling parameter 
$(k_F a_F)^{-1}$. The results of the present theory (t-matrix-I) and of its 
version without the inclusion of the self-energy shift 
$\Sigma_0$ (t-matrix-II) 
are compared with the BCS mean field (BCS), the Fixed-node QMC data from 
Ref.~\onlinecite{FNQMC} (FNQMC), the Galitskii's expression for the dilute 
Fermi gas (Galitskii), and the asymptotic expression for strong coupling using
the result $a_B=0.6 a_F$.} 
\end{center}
\end{figure} 
Figure 1 shows the comparison for the chemical potential at zero temperature,
as obtained by our calculation\cite{PPS-PRB-04} and by the Fixed Node QMC 
(FNQMC) calculations 
of Ref.~\onlinecite{FNQMC}. As discussed in 
Ref.~\onlinecite{PPS-PRB-04}, on the weak-coupling side we find it appropriate 
to introduce a constant shift $\Sigma_0$ in the bare Green's function entering 
the self-energy. This shift needs to be included only for coupling values 
$(k_F a_F)^{-1}\le -0.5$, such that the self-energy can be considered to be 
approximatively constant. The curve obtained by this procedure is reported
in Fig.~1  with the label t-matrix-I and corresponds to the data 
reported in Fig.~6 of Ref.~\onlinecite{PPS-PRB-04}. For completeness, we also 
report in Fig.~1 the curve obtained without the inclusion of the self-energy
shift $\Sigma_0$ [with the label t-matrix-II] (by definition, the two curves 
I and II coincide when
 $(k_F a_F)^{-1}\ge -0.5$).
Our results are in excellent agreement with the FMQMC data in 
the range $-0.5\lesssim(k_F a_F)^{-1}\lesssim 0.5$ spanning the crossover 
region. 

For couplings $(k_F a_F)^{-1}\le -0.5$, the FNQMC results are extremely close
to both our curves, lying just in between them.
In the weak-coupling region $(k_F a_F)^{-1}\lesssim -2$, 
our curves (as well as the FNQMC data) approach the asymptotic expression by 
Galitskii~\cite{Galitskii} for 
the chemical potential of a dilute Fermi gas.
The BCS mean field (also reported in Fig.~1) misses instead the Galitskii 
correction to the non-interacting chemical potential.
More specifically, we have verified that our theory with the inclusion
of the self-energy shift $\Sigma_0$ [t-matrix-I] recovers the complete 
Galitskii's expression 
$\mu/\epsilon_F=1+\frac{4}{3\pi} k_F a_F + \frac{4}{15\pi^2}(11-2\ln 2)
(k_F a_F)^2$ including the second-order correction in $k_F a_F$.
The curve for the chemical potential obtained without the inclusion of the 
shift $\Sigma_0$ recovers instead only the leading order
correction linear in $k_F a_F$. [It can also be shown that neglecting the 
shift
$\Sigma_0$ introduces a spurious additional term 
$\frac{3}{2}(\frac{4}{3\pi} k_F a_F)^2$ to the second-order correction in 
the Galitskii's expression.] 

\begin{figure}
\begin{center}
\includegraphics*[width=8.5cm]{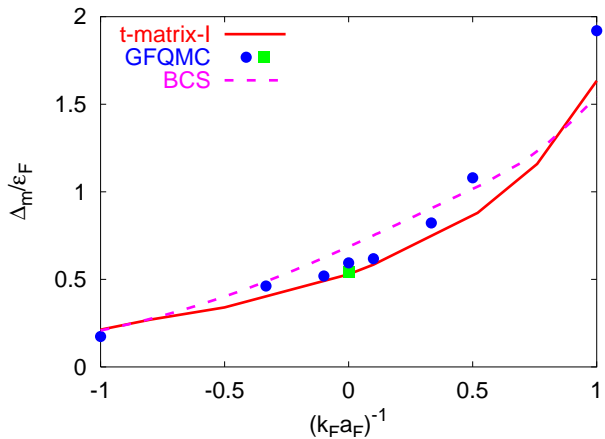}
\caption{Excitation gap $\Delta_{{\rm m}}$ at zero temperature vs the
coupling parameter $(k_F a_F)^{-1}$. The results of the present theory 
(t-matrix-I) are compared with the Green's function QMC data of 
Refs.~\onlinecite{carlson03} and ~\onlinecite{chang04} (GFQMC) as well as 
with the BCS mean field (BCS).}  
\end{center}
\end{figure} 

On the strong-coupling side, for coupling values $(k_F a_F)^{-1}\gtrsim 0.5$ 
our results deviate somewhat from the FNQMC data. This discrepancy is due to 
the fact that in our approach the boson-boson scattering is treated at the 
level of the 
Born approximation, corresponding to the value $a_B= 2 a_F$ of the bosonic 
scattering length $a_B$. 
The importance of including the correct value of the bosonic scattering length
($a_B= 0.6 a_F$, as calculated in Ref.~\onlinecite{petrov}) in this region 
 is clearly seen from the agreement between the 
FNQMC data and the asymptotic expression $\mu= -\epsilon_0/2 + \mu_B/2$, where 
$\epsilon_0$ is the binding energy of the 2-body problem and 
$\mu_B=4\pi n_B a_B/m_B$, with $n_B=n/2$, $m_B=2 m$, and $a_B = 0.6 a_F$. 
The asymptotic curve corresponding to the value $a_B = 2 a_F$ almost coincides 
with our curve in this region. [This curve is not reported in Fig.~1 
for overall clarity.]
It is, finally, interesting to mention that the inclusion of the next-order 
correction to the bosonic chemical potential, corresponding to 
the expression
 $\mu_B=\frac{4\pi n_B a_B}{m_B}[1+\frac{32}{3}(n_B a_B^3/\pi)]^{1/3}$ obtained
in Ref.~\onlinecite{leeyang}, would worsen appreciably the comparison between 
the QMC data and the asymptotic 
curve in the coupling region $0.2 \lesssim (k_F a_F)^{-1} \lesssim 2$. 
The inclusion of this next-order term improves the comparison only in the 
truly asymptotic regime for $(k_F a_F)^{-1} \gtrsim 2$ (not reported in the 
figure), where the next-order correction to the bosonic chemical potential is, 
however, already quite small.
This finding could (at least partially) explain the absence of 
beyond-mean-field corrections on the bosonic side of the 
BCS-BEC crossover, recently reported in experiments 
with ultracold Fermi gases \cite{grimm04}.

Quite generally, any theory of the BCS-BEC crossover connects the equation 
for the chemical potential $\mu$ to the equation 
for the gap (order) parameter $\Delta$ in the superfluid phase.  
The latter  quantity is not directly accessible to the QMC simulations of 
Refs.~\onlinecite{chang04} and \onlinecite{FNQMC}. In 
Ref.~\onlinecite{chang04}, however, the even-odd staggering of the 
ground-state energy for a system with a finite
number of particles was exploited to calculate the single-particle excitation
gap $\Delta_{{\rm m}}$. In a BCS-like framework (and for a sufficiently large 
number of particles) the gap $\Delta_{{\rm m}}$ is expected to coincide with
 the gap (order) parameter $\Delta$ when $\mu$ is positive and with 
the quantity $(\Delta^2+\mu^2)^{1/2}$ when  $\mu$ is negative. For a given 
coupling, this gap occurs at the wave vector $|{\bf k}|=\sqrt{2 \mu}$ for 
positive $\mu$ and at ${\bf k}=0$ for negative $\mu$. 
These BCS-like results are not expected to hold 
exactly away from weak coupling. 
The calculations presented in Ref.~\onlinecite{PPS-PRB-04}, nevertheless, show
 that the
identification of the single-particle excitation gap $\Delta_{{\rm m}}$ with 
$\Delta$ for $\mu>0$
and with $(\Delta^2+\mu^2)^{1/2}$ for $\mu<0$ works fairly well for all 
couplings of interest.
In particular, in Fig.~14 of Ref.~\onlinecite{PPS-PRB-04} this
definition of the excitation gap $\Delta_{{\rm m}}$ was compared with the 
results 
obtained from an accurate analysis of the
single-particle spectral function $A({\bf k},\omega)$, showing
that the two definitions are in good agreement with each other 
over a wide coupling range.

In Fig.~2 we compare $\Delta_{{\rm m}}$, as obtained from our results for
$\Delta$ and $\mu$, with the QMC data of Ref.~\onlinecite{chang04}. The BCS
mean field results are also reported for completeness. For the coupling value
$(k_F a_F)^{-1}=0$ a single QMC datum previously available from 
Ref.~\onlinecite{carlson03} is also reported in the figure (full square). 
Even for the excitation gap, our results appear to be in remarkable 
agreement with QMC 
data in the crossover region $-1 \lesssim (k_F a_F)^{-1} \lesssim 0.4$. At
larger couplings, the QMC results start instead to deviate from our 
results, the discrepancy being mainly due to the finite range of 
the interaction potential used in the QMC calculations. In strong coupling, 
both our excitation gap and that calculated from QMC simulations tend, in 
fact, to half the value of the binding energy $\epsilon_0$ of the two-body 
problem.
The binding energies for the contact potential and for the finite-range 
potential
used in Ref.~\onlinecite{chang04} are close to each other only in a narrow
range about $(k_F a_F)^{-1}=0$. At the coupling value 
$(k_F a_F)^{-1}=1$, the binding energy for the finite-range potential of 
Ref.~\onlinecite{chang04} is already larger by about 40\% than 
the contact-potential binding energy. This difference is responsible 
for the discrepancy between our values and the QMC data of 
Ref.~\onlinecite{chang04} on the strong-coupling side, where the excitation 
gap is controlled by the binding energy
of the two-body problem. 
 
In conclusion, the theory of Ref.~\onlinecite{PPS-PRB-04} for the BCS-BEC 
crossover in the broken-symmetry phase has been shown to compare extremely 
well with recent QMC data at zero temperature, especially in the 
intermediate-coupling (crossover) region which is the most interesting one 
both theoretically and experimentally. This agreement suggests that the choice
of the fermionic self-energy made in Ref.~\onlinecite{PPS-PRB-04} captures the 
essential physics of the problem,
as soon as the fermionic degrees of freedom get progressively quenched while 
forming composite bosons.

\acknowledgments
We thank S. Giorgini for discussions and for providing us with the data file 
of Fig.~2  of Ref.~\onlinecite{FNQMC}.


\begin{thebibliography}{99}
 
\bibitem{exp-BCS-BEC} M. Bartenstein, A. Altmeyer, S. Riedl, S. Jochim, C. 
Chin, J. Hecker Denschlag, and R. Grimm, Phys. Rev. Lett. {\bf 92}, 120401 
(2004); C.A. Regal, M. Greiner, and D.S. Jin, Phys. Rev. Lett. {\bf 92}, 040403
(2004); M.W. Zwierlein, C.A. Stan, C.H. Schunck, S.M.F. Raupach, A.J. Kerman, 
and W. Ketterle, Phys. Rev. Lett. {\bf 92}, 120403 (2004);
T. Bourdel, L. Khaykovich, J. Cubizolles, J. Zhang, F. Chevy, M. Teichmann, 
L. Tarruell, S.J.J.M.F. Kokkelmans, and C. Salomon, 
Phys. Rev. Lett. {\bf 93}, 050401 (2004).                      
\bibitem{PPS-PRB-04} P. Pieri, L. Pisani, and G.C. Strinati,
Phys. Rev. B {\bf 70}, 094508 (2004).
\bibitem{Perali02} A. Perali, P. Pieri, G.C. Strinati, and C. Castellani, 
Phys. Rev. B {\bf 66}, 024510 (2002).

\bibitem{Leggett} A.J. Leggett, in \emph{Modern Trends in the Theory of 
Condensed Matter\/}, edited by  A. Pekalski and R. Przystawa, Lecture 
Notes in Physics Vol.115 (Springer-Verlag, Berlin, 1980), p.13.

\bibitem{NSR} P. Nozi\`{e}res and S. Schmitt-Rink, J. Low. Temp. Phys. 
{\bf 59}, 195 (1985).

\bibitem{Popov} V.N. Popov, \emph{Functional Integrals and Collective 
Excitations\/} (Cambridge University Press, Cambridge, 1987).

\bibitem{carlson03} J. Carlson, S.-Y. Chang, V.R. Pandharipande, and  
K.E. Schmidt, Phys. Rev. Lett. {\bf 91}, 50401 (2003).

\bibitem{PPPS-PRL-04} A. Perali, P. Pieri, L. Pisani, and G.C. Strinati, 
Phys. Rev. Lett. {\bf 92}, 220404 (2004).

 
\bibitem{chang04} S.Y. Chang, V.R. Pandharipande, J. Carlson, and K.E. Schmidt,
Phys. Rev. A {\bf 70}, 043602 (2004).

\bibitem{FNQMC} G.E. Astrakharchik, J. Boronat, J. Casulleras, and S. Giorgini,
cond-mat/0406113 (to appear in Phys. Rev. Lett.).

\bibitem{quantitative} A. Perali, P. Pieri, and G.C. Strinati, 
Phys. Rev. Lett. {\bf 93}, 100404 (2004).


\bibitem{SPS} S. Simonucci, P. Pieri, and G.C. Strinati, cond-mat/0407600.

\bibitem{Galitskii} V.M. Galitskii, Zh. Eksp. Teor. Fiz. {\bf 34}, 151 (1958) 
[Sov. Phys. JETP {\bf 7}, 104 (1958)].

\bibitem{petrov} D.S. Petrov, C. Salomon, and G.V. Shlyapnikov, 
Phys. Rev. Lett. {\bf 93}, 090404 (2004). 

\bibitem{leeyang} T.D. Lee and C.N. Yang, Phys. Rev. {\bf 105}, 1119 (1957).

                           
\bibitem{grimm04} M. Bartenstein, A. Altmeyer, S. Riedl, S. Jochim, C. 
Chin, J. Hecker Denschlag, and R. Grimm, Phys. Rev. Lett. {\bf 92}, 203201 
(2004).

\end{thebibliography}
\end{document}